\documentclass[12pt,a4paper]{article}

\title{SI, CGSG, and $c \hbar$ units: metrology
and special relativity}
\author{L.B. Okun \\
ITEP, Moscow, Russia}

\date{}

\begin{document}

\maketitle

\begin{abstract}
Invited report at the 8th Meeting of the Working Group ``Base
units and fundamental constants'' of the French Academy of
Sciences. June 14, 2004.

The text consists of seven sections:
\begin{enumerate}
\item Great importance of SI.
\item The necessity of updating SI.
\item Special relativity: four-dimensional quantities.
\item Maxwell equations in vacuum in CGSG units.
\item Maxwell equations in vacuum in SI.
\item $c \hbar$ units.
\item Systems CGSG and $c \hbar$ should be legalized.
\end{enumerate}
\end{abstract}

\section{\large Great importance of SI}

The SI units are of great importance for science, industry,
business, trade, jurisdiction, for the knowledge-based society as
a whole. SI forms the language of modern metrology with its
world-wide network of national metrological institutes. Therefore
any improvements should be introduced into SI with maximal caution
and conservatism.

\section{\large The necessity of updating SI}

On the other hand the successes of fundamental science and its
role in the knowledge-based society necessitate the evolution of
SI, whose basic elements were laid down in XIXth century. Since
the middle of the XXth century SI has been internationally
accepted as the only legal unit system to be used in textbooks for
pupils and students. According to the legal documents of SI, other
systems of units might be exceptionally allowed in research
papers, but, as was stressed by Thibauld Damour, as nobody learns
anymore how to use such systems, new generations of scientists are
fully SI dominated. In particularly they are CGSG (CGSG =
CGSGauss) ignorant and their knowledge of special relativity is
non-adequate.

\section{\large Special relativity: four-dimensional quantities}

In CGSG units Maxwell equations in vacuum are expressed in terms
of coordinate four-vector $x^i$, of four-vector potential $A_i$,
four-vector current $j_i$ and four-tensor of electromagnetic field
$F_{ik}$.

Contravariant and covariant expressions: $$ x^i = x^0, x^1, x^2,
x^3 = (ct, {\rm\bf x}) $$ $$ x_i = (ct, -{\rm\bf x}) $$ $$ A_i =
A_0, A_1, A_2, A_3 = (\varphi, -{\rm\bf A}) $$

$$ A^i = A^0, A^1, A^2, A^3 = (\varphi, +{\rm\bf A}) $$  $$ A^l =
g^{lm} A_m  $$ $$ j_i = (\rho, {\rm\bf j}) $$ $$ F_{ik} =
\frac{\partial A_k}{\partial x^i} - \frac{\partial A_i}{\partial
x^k} $$ $$ F^{lm} = \frac{\partial A^m}{\partial x_l} -
\frac{\partial A^l}{\partial x_m} $$ $$ \tilde F_{ik} =
\frac{1}{2} \varepsilon_{iklm} F^{lm} $$ where
$\varepsilon_{iklm}$ is fully antisymmetric four-tensor ($\tilde
F_{ik}$ is dual with respect to $F_{ik}$).

\section{\large Maxwell equations in vacuum in CGSG units}

In terms of $F_{ik}$ and $j_i$ the two Maxwell equations are
simple and beautiful: $$\frac{\partial F_{ik}}{\partial x^k} =
-\frac{4\pi}{c}j_i $$ $$\frac{\partial \tilde F_{ik}}{\partial
x^k} = 0 $$

They express the gist of classical electrodynamics and together
with Lorentz transformations present one of the most remarkable
manifestations of special relativity.

The difference between CGS and CGSG is that the former has three
base units (that of length, time, and mass), while in the latter a
fourth base (electromagnetic) unit is added (that of electric
charge, or current, or electric permittivity). In this respect
CGSG is similar to SI.

\section{\large Maxwell equations in vacuum in SI}

According to SI, Maxwell equations cannot be presented as two
four-dimensional equations, but as four three-dimensional
equations for four three-dimensional vectors ${\rm\bf E}, {\rm\bf
D}, {\rm\bf B}, {\rm\bf H}$ with dimensional coefficients $\mu_0 =
1/\pi 10^{-7}$ NA$^{-2}$, $\varepsilon_0 = 1/\mu_0 c^2 = 8.854 ...
\cdot 10^{-12}$ Fm$^{-1}$, which are called magnetic permeability
and electric permittivity of vacuum and have no direct physical
meaning. They acquire such meaning when vacuum is compared with
material media for which $\mu$ and $\varepsilon$ are very
important. Both $\mu_0$ and $\varepsilon_0$ originate from the
obsolete concept of ether. The SI form of Maxwell equations hides
their beauty and the genuine physical meaning which are seen so
clearly in the CGS units.

This side of SI inevitably leads to lack of understanding of
special relativity and to deterioration of general culture in the
community of physicists and engineers, which cannot be tolerated
in a knowledge-based society.

If in CGSG $\varepsilon_0$ is chosen as the unit of permittivity,
the Coulomb law and Maxwell equation in general look identical in
CGSG and CGS. But this choice does not reduce the number of base
units in CGSG from four to three. For a similar discussion
concerning $c$ see next section.

\section{\large {\bf $c \hbar$} units}

In Quantum Field Theory, combining Quantum Mechanics and Special
Relativity, the $c, \hbar$ units are widely used. In these units
the velocity of light is the unit of velocity, while quantum
$\hbar$ is the unit of action (and of angular momentum). Quite
often this system is referred to as ``$c, \hbar = 1$ units''.
However this name is a kind of theoretical ``stenography''. When
using $c$ and $\hbar$ as units one naturally can replace in
equations factors $c$ and $\hbar$ by $c/c =1$ and $\hbar/\hbar
=1$. But this does not mean that unit of speed $c$ and unit of
action $\hbar$ are equal to unity. Such statements are much worse
than ``stenography'', they are examples of confusing theoretical
``jargon''. Unit, which is a dimensional quantity, cannot be equal
to 1 or to any other dimensionless number.

The crucial role played by $c$ in relativity and its extraordinary
stability and reproducibility have led to its nominalization
(without experimental uncertainties). However neither
nominalization, nor using $c$ as the unit of speed have led to
reduction of number of units or to reducing space to time. The use
of atomic clocks for defining distances does not reduce space to
time. Astronomers have a long tradition of using light years as a
measure of large distances. But they never claimed that time
interval abolishes distance. Space and time would be ``the same''
in Euclidean world, with metric $s^2 = t^2 c^2 + {\rm\bf x}^2$,
but our world is Minkowskian  ($s^2 = t^2 c^2 - {\rm\bf x}^2$).
Hence space is space and time is time. They are not ``the same''.

\section{\large Systems CGSG and {\bf $c \hbar$} should be legalized}

The French Academy of Sciences should recommend to the
International Committee on definition of units to explicitly allow
in the legal SI documents the use of CGS based Gaussian units not
only in research papers, but also in the textbooks and other
educational texts.

Moreover it should address SUNAMCO of IUPAP with a recommendation
on necessity to have CGSG as a part of basic electrodynamical
courses. (SUNAMCO = Symbols, Units, Atomic Masses and Fundamental
Constants, IUPAP = International Union of Pure and Applied
Physics.)

Similarly $c \hbar$-system in which the unit of velocity is $c$
and the unit of action and angular momentum is $\hbar$ should be
legalized and allowed for the textbooks on fundamental physics
both theoretical and experimental.

A certain knowledge of CGSG as well as absolute units, $c
\hbar$-units, atomic and Planck units should be a legal
requirement (based on SI) on licenses for engineers, school
teachers of physics, for university and college diplomas for
physicists.

\section*{\large Acknowledgements}

I am grateful to T.~Damour, S.~Karshenboim, G.~Veneziano and
L.~Vitushkin for valuable discussions.

\end{document}